\documentclass[useAMS,usenatbib,usegraphicx]{mn2e}
\usepackage[latin1]{inputenc}
\usepackage{amsmath,amsfonts,amssymb}
\usepackage{color}
\usepackage{times}

\newcommand{\Msun}{\ensuremath{\textrm{M}_{\odot}}}

\newcommand{\kms}{km\hspace{0.25em}s$^{-1}$}
\newcommand{\ergs}{erg\hspace{0.25em}s$^{-1}$}

\newcommand{\OI}{\mbox{O\hspace{0.25em}{\sc i}}}

\newcommand{\CI}{\mbox{C\hspace{0.25em}{\sc i}}}

\newcommand{\NaI}{\mbox{Na\hspace{0.25em}{\sc i}}}

\newcommand{\SiII}{\mbox{Si\hspace{0.25em}{\sc ii}}}

\newcommand{\CaII}{\mbox{Ca\hspace{0.25em}{\sc ii}}}

\newcommand{\FeI}{\mbox{Fe\hspace{0.25em}{\sc i}}}
\newcommand{\FeII}{\mbox{Fe\hspace{0.25em}{\sc ii}}}
\newcommand{\FeIII}{\mbox{Fe\hspace{0.25em}{\sc iii}}}

\newcommand{\CoIII}{\mbox{Co\hspace{0.25em}{\sc iii}}}

\newcommand{\Fefs}{$^{56}$Fe}

\newcommand{\Cofs}{$^{56}$Co}
\newcommand{\Nifs}{$^{56}$Ni}

\newcommand{\MCh}{M$_{\textrm{Ch}}$}

\newcommand{\aap}{A\&A}

\newcommand{\apj}{ApJ}
\newcommand{\apjs}{ApJS}
\newcommand{\apjl}{ApJ}
\newcommand{\aj}{AJ}

\newcommand{\mnras}{MNRAS}

\newcommand{\nat}{Nature}

\newcommand{\physrep}{Phys. Rep.}

\newcommand{\eg}{e.g.,\ }
\newcommand{\ie}{i.e.,\ }

\def\gsim{\mathrel{\rlap{\lower 4pt \hbox{\hskip 1pt $\sim$}}\raise 1pt \hbox {$>$}}}
\def\lsim{\mathrel{\rlap{\lower 4pt \hbox{\hskip 1pt $\sim$}}\raise 1pt \hbox {$<$}}}
\def\gtaprx {\lower .1ex\hbox{\rlap{\raise .6ex\hbox{\hskip .3ex
	{\ifmmode{\scriptscriptstyle >}\else
		{$\scriptscriptstyle >$}\fi}}}
	\kern -.4ex{\ifmmode{\scriptscriptstyle \sim}\else
		{$\scriptscriptstyle\sim$}\fi}}}
\def\ltaprx {\lower .1ex\hbox{\rlap{\raise .6ex\hbox{\hskip .3ex
	{\ifmmode{\scriptscriptstyle <}\else
		{$\scriptscriptstyle <$}\fi}}}
	\kern -.4ex{\ifmmode{\scriptscriptstyle \sim}\else
		{$\scriptscriptstyle\sim$}\fi}}}

\begin{document}

\title[A low central oxygen mass in the SN\,Ia 2010lp]
 {A very low central oxygen mass in the peculiar type Ia SN\,2010lp: further diversity at the low-luminosity end of SNe Ia}
\author[P. A.~Mazzali et al.]{P. A. Mazzali$^{1,2}$
\thanks{E-mail: P.Mazzali@ljmu.ac.uk}, 
S.~Benetti$^3$, M.~Stritzinger$^4$, C.~Ashall$^5$\\
\\
  $^1$Astrophysics Research Institute, Liverpool John Moores University, IC2, Liverpool Science Park, 146 Brownlow Hill, Liverpool L3 5RF, UK\\
  $^2$Max-Planck-Institut f\"ur Astrophysik, Karl-Schwarzschild Str. 1, 
  	D-85748 Garching, Germany\\
  $^3$INAF-Osservatorio Astronomico di Padova, vicolo dell'Osservatorio 5, Padova, Italy \\
  $^4$Department of Physics and Astronomy, Aarhus University, Ny Munkegata 120, DK-8000 Aarhus C, Denmark\\    %\href{https://orcid.org/0000-0002-5571-1833}{\includegraphics[scale=0.5]{figures/ORCID%iD_icon16x16.eps}} \\
  $^5$Institute for Astronomy, University of Hawai'i at Manoa, 2680 Woodlawn Dr., Hawai'i, HI 96822, USA
\\}

\date{Accepted ... Received ...; in original form ...}
\pubyear{2022}
\volume{}
\pagerange{}

\maketitle
  
\begin{abstract}
A nebular spectrum of the peculiar, low-luminosity type Ia supernova 2010lp is modelled in order to estimate the composition of the inner ejecta and to illuminate the nature of this event.  Despite having a normally declining light curve, SN\,2010lp was similar spectroscopically to SN\,1991bg at early times. However, it showed a very unusual double-peaked [\OI]\,$\lambda\lambda\,6300,6363$ emission at late times (Taubenberger et al.\ 2013). Modelling of the nebular spectrum suggests that a very small amount of oxygen ($\sim 0.05$\,\Msun), expanding at very low speed ($\lsim 2000$\,\kms) is sufficient to reproduce the observed emission. The rest of the nebula is not too dissimilar from SN\,1991bg, except that SN\,2010lp is slightly more luminous. The double-peaked [\OI] emission suggests that SN\,2010lp may be consistent with the merger or collision of two low-mass white dwarfs. The low end of the SN\,Ia luminosity sequence is clearly populated by diverse events, where different channels may contribute.
\end{abstract}

\begin{keywords}
supernovae: general -- supernovae: individual (SN\,2010lp) -- 
techniques: spectroscopic -- radiative transfer
\end{keywords}

\section{Introduction}
\label{sec:introduction}

While there is general consensus that type Ia supernovae (SNe\,Ia) result from 
the explosion of carbon-oxygen white dwarfs in binary systems, there is no
agreement as to the detailed properties of the progenitor system or how the
white dwarf is brought to explosion. This is a particularly interesting question
given that SNe\,Ia are used as cosmological standardizable candles
\citep{phillips93} and have heralded dark energy \citep{riess98,perlmutter99}. 

The two main contenders are the single-degenerate (SD), and the
double-degenerate (DD) scenarios. In the SD scenario, a CO white dwarf accretes
material from a non-degenerate companion. If the white dwarf has a mass close to
the Chandrasekhar limit ($\approx 1.38$\,\Msun, hereafter referred to as \MCh),
accretion causes the central temperature to rise until oxygen burning is
triggered \citep[\eg][]{whelaniben73,nom84w7}. If on the other hand the white
dwarf has a mass significantly below \MCh, a detonation can be triggered on the
surface of the white dwarf if the accreting material is He. Under specific
circumstances a shock wave can then propagate to the centre of the white dwarf
and detonate it (an ``edge-lit'' ``double detonation'') \citep{livne95}.  In the
DD scenario, two white dwarfs in a close orbit around each other merge after
radiating gravity waves \citep{webbink84,ibentutukov84}, and the resulting
massive white dwarf explodes, leaving no remnant, if it avoids accretion-induced
collapse \citep{nomoto91}.  Explosion is more likely if the merging occurs
``violently'' \citep{pakmor10}. Recently, two more scenarios have been proposed.
The head-on collision of two white dwarfs, possibly aided by a third body in a
triple system, is a variation of the DD scenario which may lead to explosion and
the creation of two distinct ejecta \citep{rosswog09,kushnir13}. This scenario
has specific observational consequences \citep{dong15}. Finally, the explosion
of a white dwarf engulfed in a common envelope by the outer layers of an
asymptotic giant branch (AGB) star and merging with the AGB star's degenerate
core, the ``core-degenerate'' scenario, has also been proposed as a channel to
produce SNe\,Ia \citep[\eg][]{livioriess2003,kashisoker2011,ashall21}. 
\citet{liviomazz2018} review possible progenitor/exposion channels.

While it is difficult to imagine how all these physically motivated scenarios,
some of which are quite aspherical in nature, might possibly coexist at all SN
luminosities and yield events that follow quite closely the relation between SN
luminosity and light curve width \citep{yungelson15}, it is also not
unreasonable to expect that they all exist, perhaps in some corner of parameter
space, while the main relation may be due to a single, dominant channel.   

The question we may ask is then, is there evidence in the data for systematic
differences along (or across) the luminosity sequence, which could hint to the
presesence, or even dominance, of a particular channel, at least in specific
luminosity/light-curve width ranges. This search does not appear to yield clear
results if it is performed based on early-time data. The temperature in the
line-forming region is a smooth function of luminosity
\citep{nugent95,hachinger08}. The luminosity distribution of SNe\,Ia, however,
seems to show two peaks: a main one, where all normal and moderately peculiar
SNe lie \citep[including the ``hot'' 1991T class][]{filippenko92b,phillips92},
and a secondary, low-luminosity peak, which includes all rapidly declining SNe,
most notably the ``cool'' 1991bg class \citep{filippenko92a,leib93,tura1996}.
These two peaks appear to be quite distinct, with few transitional events
straddling in between \citep{ashall16b}. The low-luminosity group is not only
sub-dominant in number. It also shows a strong preference for passive galaxies,
suggesting that low-luminosity SNe\,Ia originate from older progenitor systems.
Simple evolutionary considerations suggest that scenarios involving DD systems
may be on average older than SD ones. Although exceptions are possible that can
alter this picture somewhat, the evidence for this remains circumstantial and 
limited. 

We have shown in previous work that the late-time (nebular) spectra of SNe\,Ia,
which probe the inner ejecta, may more easily yield information about the
physics of the explosion and thus, more or less directly, those of the
progenitors \citep[\eg][]{mazz91bgneb}. One striking feature distinguishing
low-luminosity SNe\,Ia from normal ones at late times is the dominance of
[\FeIII] emission at the lowest velocities. These lines are also present in
``normal'' SNe\,Ia, but they typically coexist with [\FeII] lines. This
unexpected property was observed in SN\,1991bg, where [\FeIII] lines dominate at
epochs beyond 200 days and velocities below 3000\,\kms
\citep{tura1996,mazz91bgneb}, and similarly in the 91bg-like SN\,1999by
\citep{silverman12}. They are also strong in SN\,2003hv
\citep{leloud05,mazzali_03hv} and in SN\,2007on, where a complex spectrum
requires two components with different - opposite - line-of-sight velocities
\citep{mazzali_07on}. At the low densities of the late-time SN ejecta (at epochs
exceeding 300 days for normal SNe, but as early as 200 days for sub-luminous
ones), a high ionization is less the result of a high temperature than it is of
low density, which suppresses recombination more efficiently than high
temperature favours it. Therefore, if low-luminosity SNe\,Ia show a higher
degree of ionization at late times, this is likely to indicate a low central
density. These SNe are also typically characterised by lower velocity ejecta at
early times \citep{benetti05}. The combination of these factors, as well as the
early transition to the nebular regime, suggests that low-luminosity SNe\,Ia are
the explosion of progenitors of lower mass than normal SNe\,Ia. Their
progenitors are likely to be white dwarfs that did not reach \MCh\ when they
exploded. Such white dwarfs can only be exploded by compression, in violent
events such as mergers or collisions of two white dwarfs. 

There are not many low-luminosity SNe\,Ia for which nebular spectra are
available. This is in part due to their faintness, but also to the relative
scarceness of these events. It is therefore interesting to analyse data that are
available, in order to test different explosion scenarios. This is particularly
true when the spectra show peculiarities that had not been seen before.  

One outstanding example of this is SN\,2010lp. While it appeared to be similar
spectroscopically to SN\,1991bg near maximum, showing a cool spectrum and low
line velocities \citep[][see also Sect. \ref{sec:redd} below]{prietomorr11}, it
was not very sub-luminous. An estimated optical peak luminosity of $\sim 2.5
\times 10^{43}$\,\ergs\ \citep{kromer13} is comparable to that of
low-luminosity, transitional SNe\,Ia such as 1986G \citep{ashall16a}. The
decline rate of the light curve after maximum,  
\citep[$\Delta m_{15}(B) \approx 1.25$\,mag,][]{kromer13}, however, was similar to that of normal SNe\,Ia. In the
nebular phase, SN\,2010lp was again similar spectroscopically to SN\,1991bg,
except that, instead of showing narrow [\FeIII] lines at low velocities, it
displayed [\OI] 6300,6363\,\AA\ emission \citep{taub13}. Not only had these
lines not been seen before in any SN\,Ia, but they were very narrow in
SN\,2010lp, suggesting that they originate in the centre of the ejecta, and they
showed two components, one blue-shifted and the other red-shifted by similar
velocities. Double-peaked profiles had been observed only in the \NaI\,D line in
SN\,2007on \citep{dong15}, but the two components were not as distinct
\citep{mazzali_07on}. The clear separation of the two components of the [\OI]
line in SN\,2010lp suggests that the emission originates in two separate,
low-velocity blobs of material, as both spherically symmetric options, a shell
and a disc configuration, would not create such sharp separation between the two
peaks. A transition to an [\FeIII]-dominated phase was not observed in
SN\,2010lp, indicating that at low velocities iron was absent, or if it was
present either the density was not very low or that the presence of oxygen led
to a higher electron density $n_e$, making recombination easier. The late-time
spectrum of SN\,2010lp was taken at an epoch of 264 days after maximum.
\citet[][Fig.\ 4]{taub13} compare it to the spectrum of SN\,1991bg obtained
$\sim 200$ days after peak, showing the impressive difference between them
despite the underlying similarity.

\citet{kromer13} compared an early-time spectrum and the light curve of
SN\,2010lp with their simulation of a violent merger of two CO white dwarfs of
mass 0.90 and 0.76\,\Msun, and showed good agreement, which supports the merger
scenario for this peculiar SN\,Ia. In this particular type of merger,
\citet{kromer13} found that the centre of the ejecta are dominated by unburned
oxygen coming from the disrupted secondary. This is at least one step towards
what is observed in SN\,2010lp, but the central oxygen has an almost spherically
symmetric distribution, which is unlikely to produce two distinct components of
the [\OI] emission in the nebular phase. The analysis of \citet{kromer13} did
not extend to the nebular epoch. Further study is therefore warranted. A merger
scenario had been proposed for SN\,1991bg itself \citep{pakmor11,mazz91bgneb},
suggesting that the merger channel represents a significant contribution to
low-luminosity SNe\,Ia. 

As the analysis of the nebular spectrum in \citet{taub13} was mostly
qualitative, we aim to characterise the inner ejecta more quantitatively. In
particular, we determine the abundance and distribution of the central oxygen. 
To this end we use our non-local thermodynamic equilibrium (NLTE) SN nebular
code, which is briefly described in Section \ref{sec:method}. Before that, we
present a low resolution spectrum of SN\,2010lp obtained at Gemini that had not
been analysed before. As the spectrum shows interstellar \NaI\,D absorptions
from both the Galaxy and the host \citep[which has redshift
$z=0.010$,][]{taub13}, it is very useful in order independently to determine the
reddening to SN\,2010lp (Sect. \ref{sec:redd}), which turns out to be smaller
than previously estimated. In Sect. \ref{sec:results} our possible solution for
the nebular spectrum of SN\,2010lp is presented, and in Sect. \ref{sec:disc} our
results and their implications are discussed. Sect. \ref{sec:concl} concludes
the paper.

\begin{figure*} 
\includegraphics[width=139mm,angle=-90]{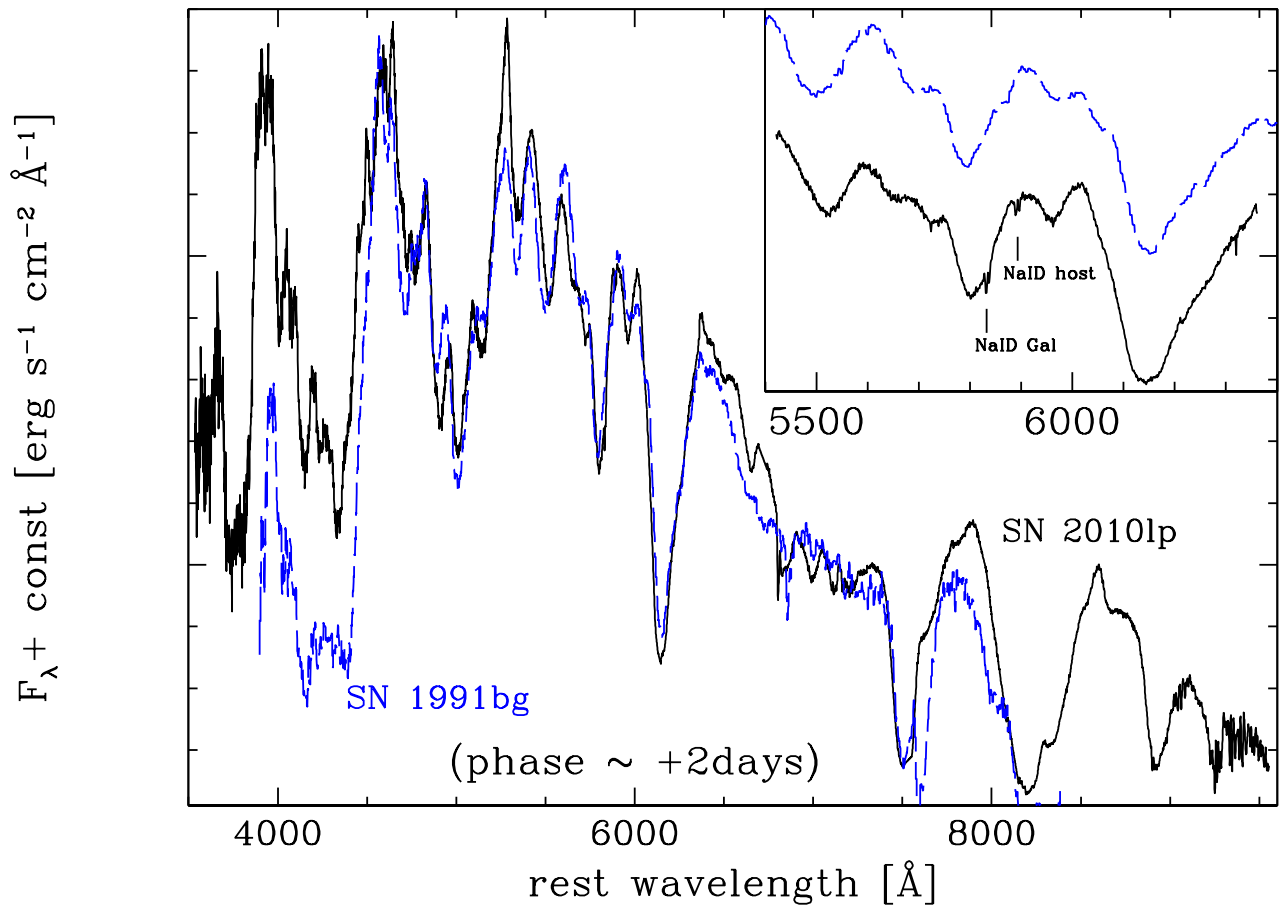}
\caption{A comparison of the spectrum of SN\,2010lp obtained with Gemini at a 
phase $\approx 2$ days after maximum light (black solid line) with the coeval 
spectrum of SN\,1991bg (blue dashed line). Both spectra have been corrected for 
reddening (E(B--V) = 0.05 mag for SN\,1991bg and 0.14 mag for SN\,2010lp) and 
redshift. The galactic and host interstellar \NaI\,D doublets visible in the 
spectrum of SN\,2010lp are marked.}
\label{fig:10lp_spec}
\end{figure*}

\section{Gemini-N spectrum and reddening}
\label{sec:redd}

One of the most uncertain parameters that may affect the modelling of SN\,2010lp
is reddening. 

Here we present and analyse an early, unpublished spectrum of SN\,2010lp
obtained with the 8.1-m Gemini-North telescope equipped with the Multi-Object
Spectrograph GMOS-N. The data were obtained using the B600+G5307 disperser with
a a central wavelength of 500\,nm and the R400+G5305 disperser with a central
wavelength of 750\,nm.  Observations were performed on Jan. 10, 2011
(MJD(avg)=55571.291; i.e., $\sim 2$ days after the time of maximum assumed by
\citet{taub13}. They consisted of 600 seconds of integration time, and cover the
spectral range 3570 to 9625\,\AA. The data were reduced using the \texttt{gemini
gmos} package contained within the \texttt{IRAF} environment. The extracted 1-D
spectra of SN\,2010lp were flux-calibrated using a sensitivity function computed
from archival observations of the standard star BD+28\:4211 obtained on
MJD\,55534.27. The combined, flux-calibrated spectrum is shown in Fig.\,1, 
where it is compared with a coeval spectrum of SN\,1991bg dereddened with
$E(B-V)_{tot} = 0.05$ mag assuming a standard extinction law \citep{card89}.

Interestingly, the spectrum of SN\,2010lp shows clear \NaI\,D interstellar
absorption lines. A Galactic component has equivalent width EW(\NaI\,D)$_{\rm
Gal}=0.69$\,\AA, while the component at the redshift of the host galaxy
($z=0.010$) has EW(\NaI\,D)$_{\rm host}=0.35$\,\AA. Since Galactic reddening is
E(B--V)$_{\rm Gal}=0.094$\,mag \citep{shafly11} using the same relation between
EW(\NaI\,D) and E(B--V) as in the Galaxy (E(B--V) / EW(\NaI\,D)$ =
0.136$\,mag\,\AA$^{-1}$), we obtain E(B--V)$_{\rm host}=0.048$\,mag. This is
intermediate between the values derived using two empirical relations between
EW(\NaI\,D) and reddening. One, presented by \citet{tur2003}, gives E(B-V)$_{\rm
host}=0.056$\,mag, while the other, proposed by \citet{poz2012}, yields
E(B--V)$_{\rm host}=0.036$\,mag. These relations are subject to a large
uncertainty \citep{phillips2013}, such that we may estimate E(B--V)$_{\rm
host}=0.048 \pm 0.012$\,mag. The total reddening towards SN\,2010lp is then
E(B-V)$_{\rm tot}=0.14 \pm 0.01$\,mag. When the spectrum of SN\,2010lp is
dereddened by this amount, both the continuum and line features look very
similar to the spectrum of SN\,1991bg.

The velocity of the \SiII\,6355\,\AA\ absorption in this near-maximum spectrum
is $\sim 9800$\,\kms. This low expansion velocity is typical of the Faint SN\,Ia
subclass as defined in \citet{benetti05}, whose prototype is actually
SN\,1991bg. This confirms the similarity of the two SNe.

\section{Method}
\label{sec:method}

Synthetic nebular spectra were computed using our NLTE code. The code is based
on the assumptions outlined in \citet{axelrod80}. The gas in the SN nebula is
assumed to be heated by collisions with the high-energy particles generated in
the thermalization process of the gamma-rays and positrons emitted in the decay
chain \Nifs\ $\rightarrow$ \Cofs\ $\rightarrow$ \Fefs, and it cools via the
emission of (mostly) forbidden lines. Some strong permitted transitions are also
considered.

The radioactive decay of \Nifs\ and \Cofs\ produces both gamma-rays and
positrons, which deposit their energy in the SN ejecta and thus power the
SN light curve. Gamma-rays carry most of the radioactive energy released by the decay. Positrons are responsible for only $\approx 3.6$\% of it. Deposition is computed using a Montecarlo method, as outlined in \citet{cappellaro97} and \citet{mazzali2001a}. Constant opacities are used for both processes: $\kappa_{\gamma}=0.027$\,cm$^2$g$^{-1}$ for gamma-rays and $\kappa_{e^+}=7$\,cm$^2$g$^{-1}$ for positrons. At the times considered here the density in the ejecta is still high enough that gamma-rays deposit efficiently enough to make the dominant contribution to the energy deposition, despite the significantly lower opacity to which they are subjected.

After computing the energy deposition, the ionization and the thermal balance
are solved in NLTE \citep{ruizlaplucy92}. Ionization is assumed to be entirely due to impact with the high-energy particles produced by the deposition of the radioactive products, while photoionization is assumed to be negligible
\citep{kozmafransson98}. The rate of impact ionization and the recombination
rate are balanced for each ion to compute the degree of ionization. Level
populations are computed solving the rate equations under the assumption of
thermal balance, \ie\ equating the non-thermal heating rate and the rate of
cooling via line emission. Under the assumption that the nebula is optically
thin, radiation transport is not performed. The resulting line emissivity is
used to compute the emerging spectrum. 

The code has been used for both SNe\,Ia \citep[\eg][]{mazz91bgneb,mazzali_03hv} and SNe\,Ib/c \citep[\eg][]{mazzali06ajneb}, and it can yield a description of the inner layers of the SN ejecta. Both a one-zone and a one-dimensional version are available. The latter treats ejecta stratification in both density and abundance.

\begin{figure*} 
\includegraphics[width=139mm]{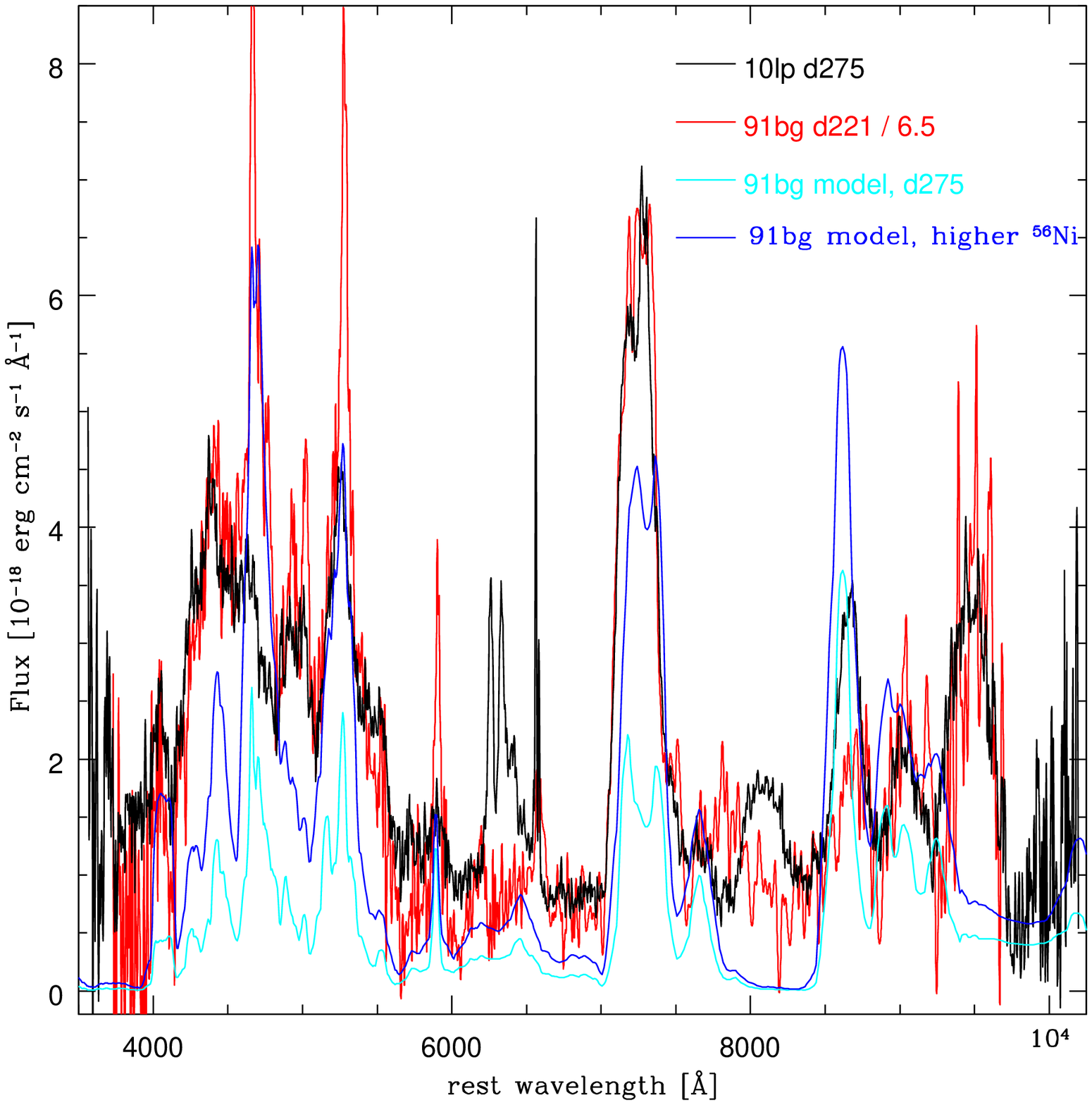}
\caption{The nebular spectrum of SN\,2010lp obtained 264 days after maximum 
\citep[][black]{taub13} compared to the nebular spectrum of SN\,1991bg (red)
obtained combining the spectrum obtained $\sim 203$ days after maximum by
\citet{tura1996}, which covers the blue side and the one obtained by R. Lopez
$\sim 199$ days after maximum and presented by \citet{taub13}, rescaled down 
in flux to match roughly the flux of SN\,2010lp. Also shown are two synthetic 
spectra. The one drawn in cyan was computed using the sub-Chandrasekhar mass
model for SN\,1991bg discussed in \citet{mazz91bgneb}, while the one drawn in 
blue is based on the same model but has a larger \Nifs\ mass 
($\approx 0.084$\,\Msun).}
\label{fig:10lp_91bg}
\end{figure*}

\section{Results}
\label{sec:results}

\citet{taub13} convincingly showed that SN\,2010lp displayed a nebular spectrum closely resembling that of SN\,1991bg about 200 days after maximum. The main difference was the absence of the narrow [\FeIII] lines. These were replaced by two narrow (full width at half maximum $\approx 1900$\,\kms) emissions which correspond to [\OI] 6300, 6363\,\AA, with a blue- and a red-shifted component. \citet{mazz91bgneb} showed that the higher ionization [\FeIII] lines in SN\,1991bg are emitted from a small volume, presumably at the centre of the ejecta, bounded by a velocity of $\sim 3000$\,\kms. A low central density was likely to be responsible for the high ionization.  

As \citet{taub13} remarked, the width of the [\OI] lines suggests that they are
also emitted at the centre of the ejecta, by the lowest-velocity material.
\citet{taub13} wondered how the same ejecta structure that produced \FeIII\ in
SN\,1991bg can yield a low ionization and neutral oxygen lines (at a later
epoch) in SN\,2010lp. We argue that the different mean molecular weight of an
oxygen-dominated gas (with mean molecular weight $\mu \approx 16$) with respect
to an iron-dominated gas ($\mu \approx 56$) would lead to a similar electron
density $n_e$ in the two gases even if on average only every other oxygen atom
released an electron in an oxygen-dominated gas, while every iron atom released
$\sim 1.5$ electrons (such that both \FeII\ and \FeIII\ are present in similar
amounts) in an iron-dominated gas.  Recombination depends on the product $n_e
n_{ion}$, where $n_{ion}$ is the number density of ions. Given that $n_{ion}$ in
a partially ionized oxygen-dominated gas is $\sim 3.5$ times as large as in an
iron-dominated gas, we may expect the ionization to be lower in the former at
similar mass density. 

As a proof of principle, we therefore started from the density/abundance profile
(the ``explosion model'') that \citet{mazz91bgneb} used for SN\,1991bg, and
replaced iron with oxygen in the innermost layers, up to a velocity of 2000\,\kms, in agreement with the width of the observed [\OI] lines. Below we present and discuss the various steps of the procedure we adopted. 

\begin{figure*} 
\includegraphics[width=139mm]{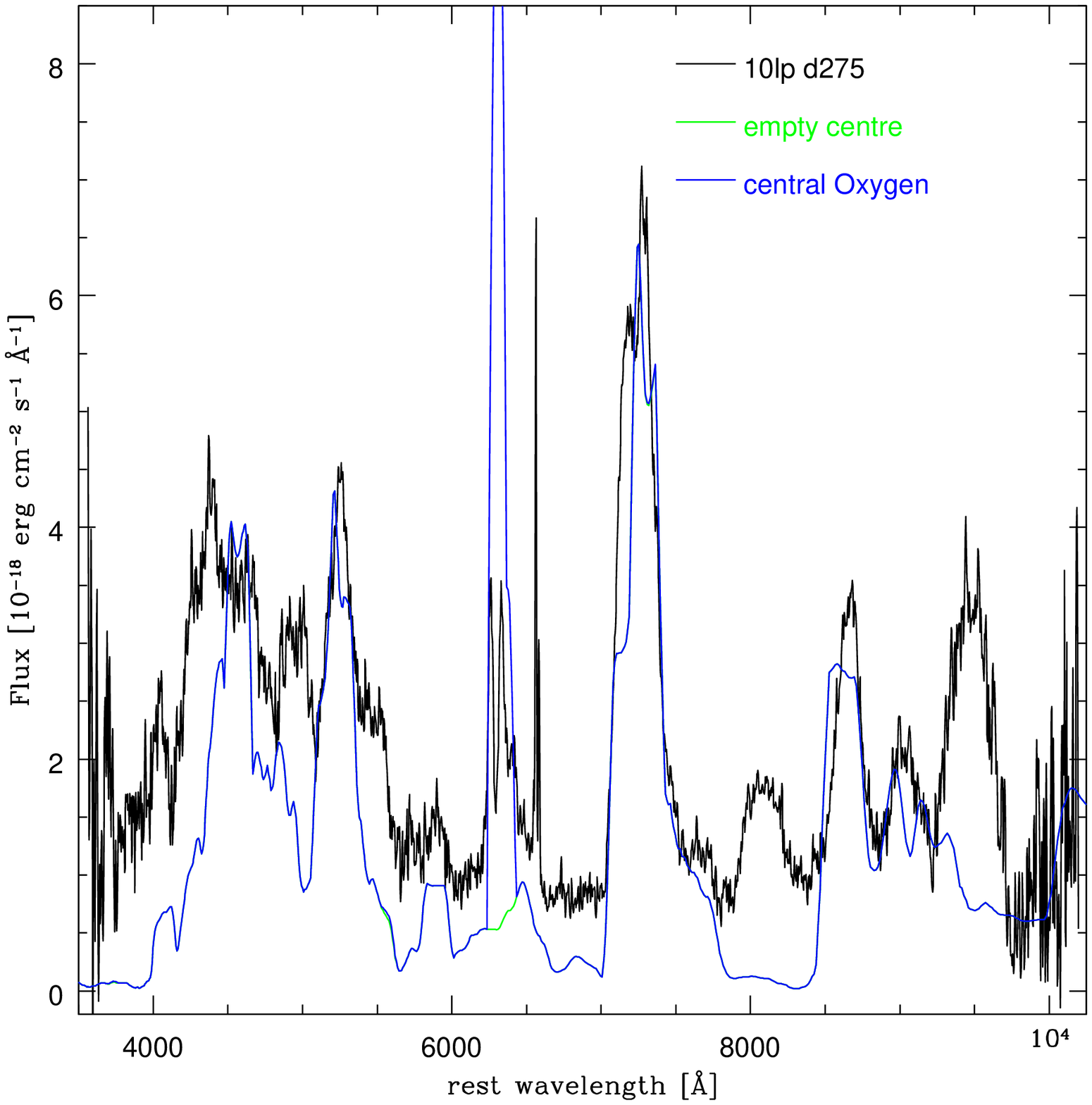}
\caption{The nebular spectrum of SN\,2010lp compared to two synthetic spectra. 
The one drawn in green (which is mostly overwritten by the other spectrum) was 
computed with the SN\,1991bg model, an increased \Nifs\ mass as shown in Fig. 
\ref{fig:10lp_91bg}, but an empty inner region (at $v < 3000$\,\kms). It shows 
a rather flat-topped [\OI] $\lambda\lambda\,6300,6363$ line (which is not 
exactly flat-topped because of the doublet nature of the emission and because 
other lines also contribute in that region). The synthetic spectrum shown in 
blue was computed using the same model except that the inner region (inside 
3000\,\kms) was filled with oxygen rather than \Nifs, and therefore it shows a 
very strong [\OI] line.}
\label{fig:10lp_empty_Ofill}
\end{figure*}

First, we compare the spectra of SNe\,2010lp and 1991bg with synthetic spectra
obtained with the model that was used to match SN\,1991bg.  Fig.
\ref{fig:10lp_91bg} shows the spectrum of SN\,2010lp obtained 264 days after
maximum \citep{taub13} and that of SN\,1991bg obtained 203 days after maximum (corresponding to an epoch $\approx 221$ days after explosion), scaled down in flux to match the spectrum of SN\,2010lp. Both spectra are shown in the rest frame and without a reddening correction. Two synthetic spectra are also shown: one is the model used for SN\,1991bg \citep{mazz91bgneb}, now computed at an epoch of 275 days, which should be a reasonable approximation for the epoch of the spectrum of SN\,2010lp, and the other is a spectrum obtained using the same density structure and epoch but with an increased \Nifs\ mass ($0.08$\,\Msun\ instead of $0.06$\,\Msun), which yields a closer match to the observed flux of SN\,2010lp (but not to its detailed features). The synthetic spectra are reddened with $E(B-V)_{\rm tot} = 0.14$, the reddening assumed for SN\,2010lp. As expected, the synthetic spectrum reproduces some features of SN\,2010lp, but shows a narrow [\FeIII] emission feature near 4700\,\AA, which is absent in SN\,2010lp. The feature near 5200\,\AA\ contains [\FeII] emission from the surrounding layers and is not completely suppressed even in SN\,2010lp.

\begin{figure*} 
\includegraphics[width=139mm]{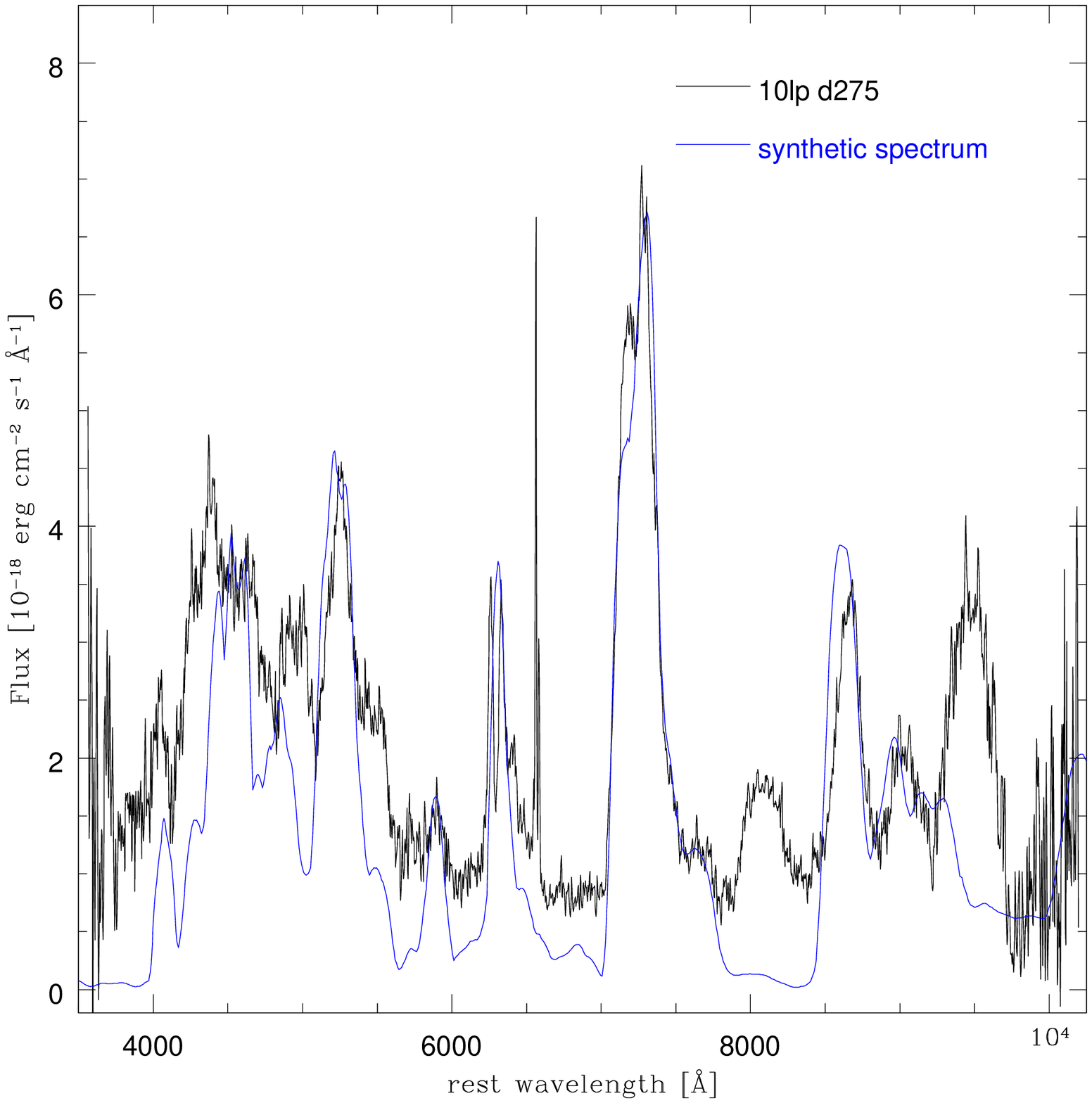}
\caption{The nebular spectrum of SN\,2010lp (black) compared to a synthetic spectrum computed using the model originally used for SN\,1991bg, with a larger \Nifs\ mass, a central oxygen-rich zone and modified abundances to yield a better match with SN\,2010lp (blue).}
\label{fig:10lp_centrO_best}
\end{figure*}

In order to reproduce the spectrum of SN\,2010lp we then performed a number of
experiments. The lack of central [\FeIII] emission and the presence of the [\OI]
lines suggest that one should test what happens to the SN\,1991bg model if the
inner iron-rich zone, inside a velocity of 3000\,\kms\ is removed and replaced
with oxygen. This is shown in Fig. \ref{fig:10lp_empty_Ofill}. Two synthetic
spectra are shown in that figure. In one the central region has been completely
removed. As expected, this spectrum does not show the narrow [\FeIII] peaks. It
is characterised by flat-top profiles in several lines (\eg\, \NaI\,D/[CoIII],
the \CaII\, IR triplet, which is however the result of a blend with [\FeII]
emission), reflecting the lack of low-velocity emission. The other synthetic
spectrum was obtained by filling the inner region with oxygen only. This
corresponds to a low-velocity oxygen mass of 0.065\,\Msun. The corresponding
synthetic spectrum shows an extremely strong [\OI] line (obviously the two peaks
are not expected to be reproduced in this one-dimensional, spherically symmetric
model). Oxygen is heated by the particles produced in radioactive decays just
outside the inner zone. The surrounding \Nifs\ zone, although not particularly
rich in \Nifs\ (0.08\,\Msun, as in the synthetic spectrum shown in Fig.
\ref{fig:10lp_91bg} as ``higher \Nifs''), is more than sufficient to excite the
upper levels of the [\OI]\,$\lambda\lambda 6300,6363$\,\AA\ transition in the
centre of the ejecta. 

Given that even a rather small amount of \Nifs\ surrounding a central,
oxygen-dominated zone is sufficient to cause much stronger [\OI] emission lines than observed, and that the mass of \Nifs\ is constrained by the overall SN luminosity, it is likely that a smaller mass of central oxygen is present in SN\,2010lp than in the model presented in Fig. \ref{fig:10lp_empty_Ofill}. If we choose not to modify the density structure that was adopetd for SN\,1991bg (the main reason for this is to avoid creating an untested explosion model as neither light curve information nor multiple spectra are available for SN\,2010lp) we need to replace some of the oxygen at low velocity with other elements. Some lines (\eg \NaI\,D/[\CoIII]) show flat-top profiles in the model with central oxygen only, which justifies this procedure. We therefore replace some of the central oxygen with a mixture containing a small amount of \Nifs, as well as the Intermediate Mass Elements (IME) that are typically abundant in the ejecta of a SN\,Ia (Si, S, Ca). Despite the additional heating provided by the centrally located \Nifs, cooling by ions other than \OI, combined with the reduced oxygen abundance, reduces the emission in the [\OI] doublet significantly. The model has a central oxygen mass (below 3000\,\kms) of 0.035\,\Msun. The \Nifs\ content inside that velocity is only 0.015\,\Msun, while IME account for 0.015\,\Msun. The synthetic spectrum, which is shown in Fig. \ref{fig:10lp_centrO_best}, offers a reasonable reproduction of the observed spectrum. We did not try to optimise it any further, as the observed [\OI] profile cannot be reproduced exactly with a one-dimensional model. The important result is that a very small mass of oxygen at low velocity is sufficient to generate emission lines comparable in strength to the observed ones in SN\,2010lp. This should impact on the possible explosion scenarios for SN\,2010lp. 

\begin{figure*} 
\includegraphics[width=139mm,angle=-90]{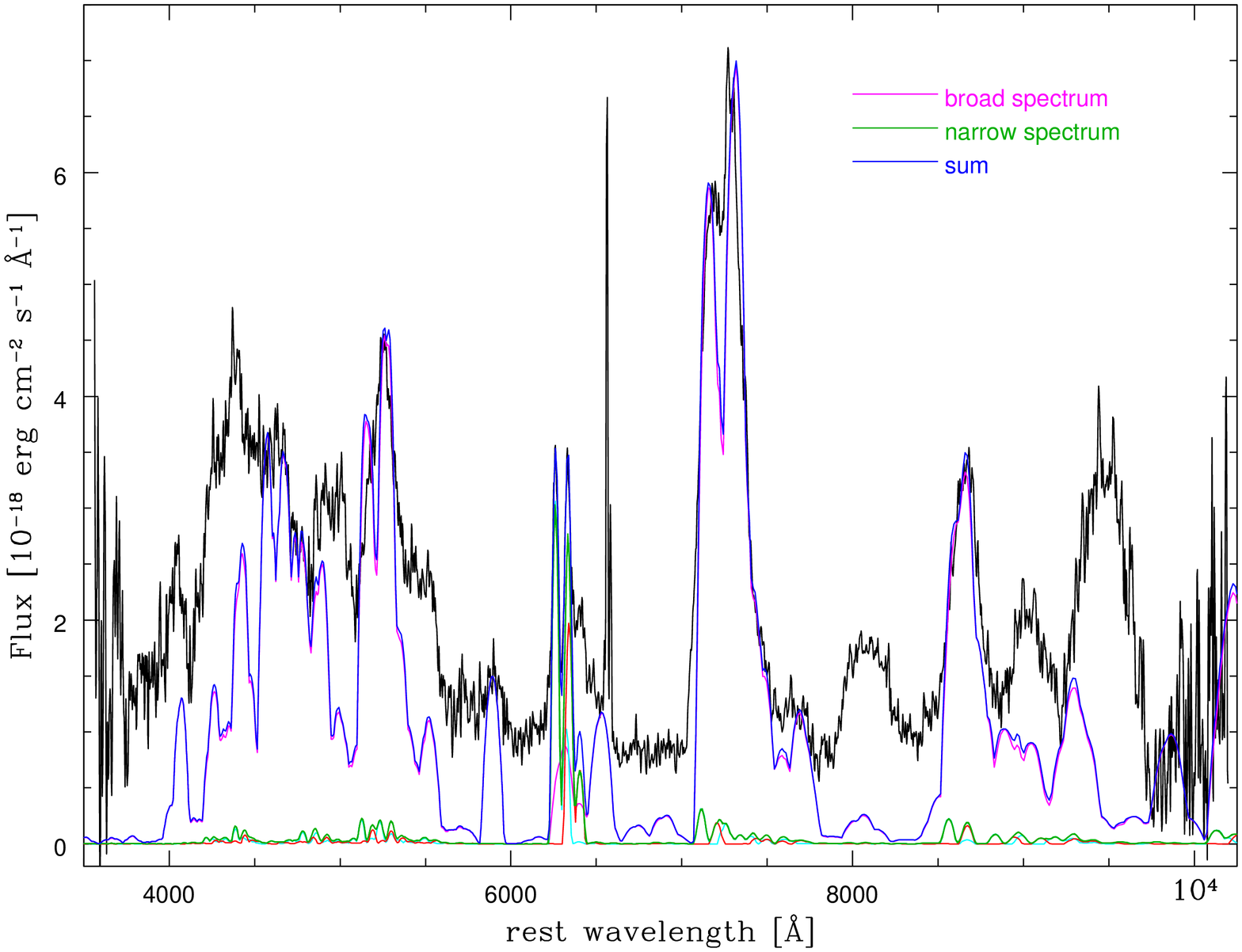}
\caption{The nebular spectrum of SN\,2010lp (black) compared to a synthetic 
spectrum obtained adding a one-zone model to reproduce the broader emissions 
(magenta) and two one-zone models that match the narrow blue- (cyan) and 
red-shifted (red) [\OI] emissions respectively. The blue/red-shift of the 
oxygen lines is $\pm 1900$\,\kms. The combined narrow-lined spectrum is drawn 
in green. The spectrum drawn in blue is the sum of the broad and narrow 
components.}
\label{fig:10lp_1z_sum}
\end{figure*}

While the estimate of the oxygen mass appears to be reasonable, in order to reproduce in detail the shape of the [\OI] emission a multi-dimensional approach is required. Such a clearly defined double-peaked [\OI] emission as is observed in SN\,2010lp cannot be obtained from a shell or an edge-on disc. In both of these cases significantly more emission at zero line-of-sight velocity would be seen, while in SN\,2010lp the two [\OI] peaks, which are blue- and red-shifted by $\sim 2000$\,\kms, respectively, are well separated. The more likely scenario is that two oxygen-rich blobs are ejected in opposite directions when the explosion occurs. As it is most likely that the actual direction of motion of the two blobs is not at a very large angle with respect to the line of sight, we may assume that the blobs have low spacial velocity ($\lsim 3000$\,\kms) and therefore remain embedded within the inner ejecta, where they can be excited by the radioactive decay products from the \Nifs\ that surrounds them and possibly coexists with them. This is a complicated scenario to simulate, so we attempt to reproduce it qualitatively adding together synthetic spectra computed as one-zone models.  

\begin{figure*} 
\includegraphics[width=139mm]{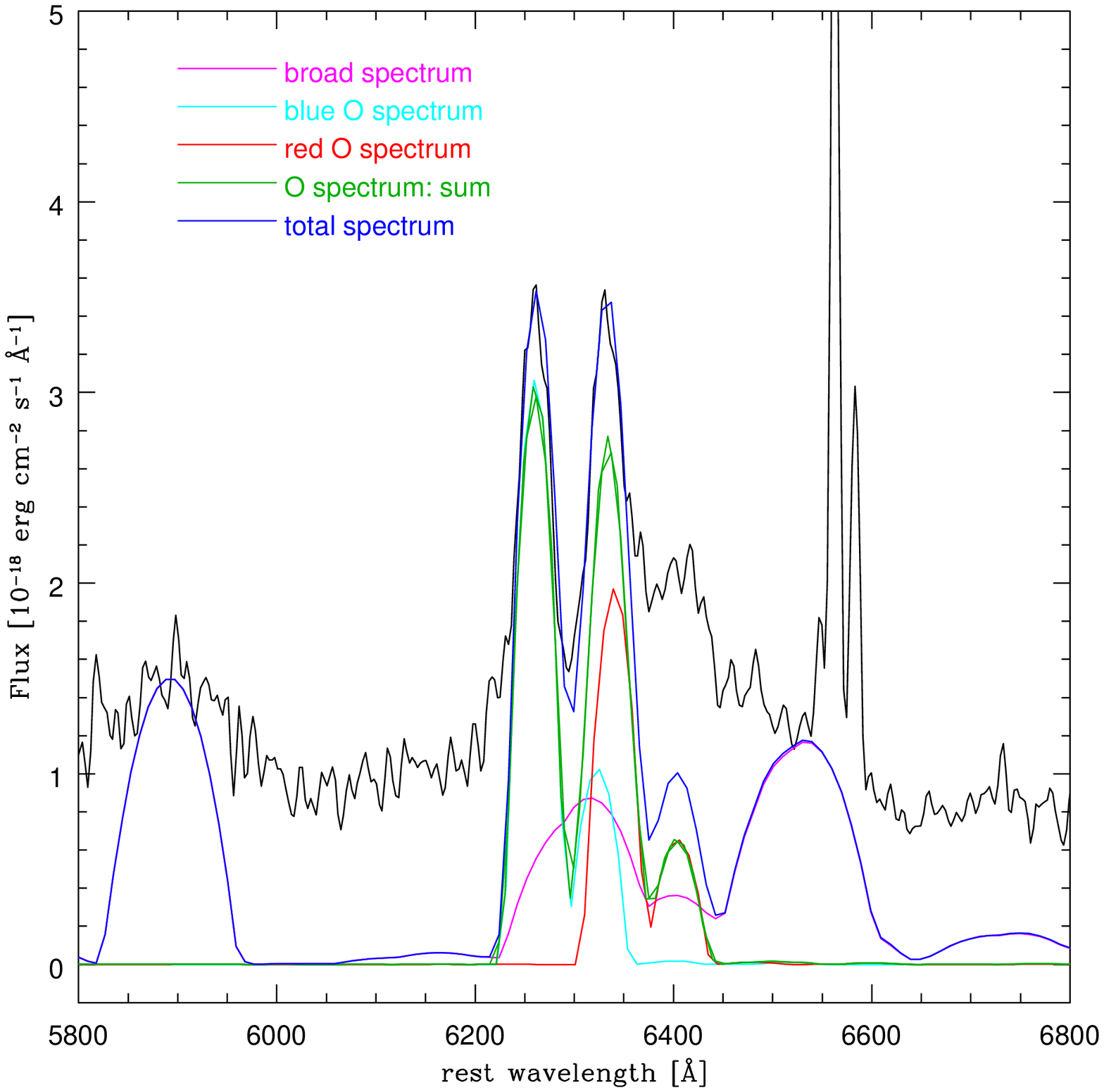}
\caption{A blow-up of Fig. \ref{fig:10lp_1z_sum} showing the [\OI] zone in more detail. The various components are as in Fig. \ref{fig:10lp_1z_sum} and are shown in the legend.}
\label{fig:10lp_1z_sum_OI_blowup}
\end{figure*}

Using a one-zone approach we produce the spectra of two oxygen-rich blobs, and add them to a one-zone simulation of the ejecta. While this is interesting as a proof-of-principle, the details of the result should be treated as an approximation at best. Two small blobs are modelled. One is blue-shifted by 1900\,\kms, the other is red-shifted by the same velocity with respect to the observer. Both blobs have boundary velocity of 1500\,\kms. The blue-shifted blob contains 0.019\,\Msun\ of oxygen,  0.002\,\Msun\ of \Nifs\  (which is used in these models to excite the gas), and small amounts of IME, for a total mass of 0.022\,\Msun. The red-shifted blob contains 0.013\,\Msun\ of oxygen and 0.0014\,\Msun\ of \Nifs, for a total mass of 0.015\,\Msun. In both cases, the \Nifs/O ratio is very small, $\lsim 0.1$. When the spectra of these two blobs are summed to the rest of the spectrum emitted by SN\,2010lp at higher velocities, a reasonable reproduction of the observed profile is obtained. The outer ejecta are treated as a single emitting zone bounded by an outer velocity of 3500\,\kms, containing a total mass of $\approx 0.25\,\Msun$ including 0.07\,\Msun\ of \Nifs\ and 0.02\,\Msun\ of oxygen. Fig. \ref{fig:10lp_1z_sum} shows the overall spectrum, while Fig. \ref{fig:10lp_1z_sum_OI_blowup} is a blow-up of the [\OI] emission region. The large mass depends on the inclusion of a significant mass of IME ($\sim 0.15\,\Msun$). The relatively low boundary velocity of the nebula leads to a split of the \CaII] 7291, 7323\,\AA\ line from the [\FeII]-dominated emission, with strongest lines at 7155 and 7172\,\AA. It also offers a better match to the emission near 8700\,\AA, which is a blend of the \CaII\,IR triplet (8498, 8542, 8662\,\AA), 
[\FeII] 8617\,\AA, and [\CI] 8727\,\AA.  

\section{Discussion}
\label{sec:disc}

While it was not the aim of this work to reproduce the nebular spectrum of
SN\,2010lp exactly, we have shown that a density structure similar to that of
SN\,1991bg, characterised by a low central density, reproduces the main features
of the observed spectrum. When the inner, \Nifs-rich region is replaced by an
oxygen-dominated mixture of lighter elements the [\OI] emission seen in
SN\,2010lp can be reproduced in overall intensity. A more exact match of the
double-peaked emission profile requires using two small blobs moving in opposite
directions at low velocity. The oxygen mass that is required to match the
emission line intensity is quite small, a total of $\approx 0.04$\,\Msun. There
is of course significant uncertainty on these values. Lacking a definitive
bolometric light curve\footnote{\citet{kromer13} present a preliminary
bolometric light curve of SN2010lp, but we do not know what reddening was used
and do not have the original photometry, which is still unpublished.} we cannot
confidently estimate the overall \Nifs\ mass, which affects the excitation of
the central oxygen at late times. The lack of a nebular spectral series makes it
difficult to determine accurately the densities in the inner layers: the
changing ionization as a function of time is very important in this respect
\citep[\eg][]{mazz91bgneb,mazzali_14jneb}.

We should note here that a few features in the red part of the spectrum are not
reproduced. In particular, emissions near 8100, 9000 and 9500\,\AA\ are weak or
absent in our synthetic spectra. The emission near 8100\,\AA\ may have
contributions from [\FeI], but the relative expected line strengths do not match
the observed profile, and other [\FeI] lines are not seen. Our synthetic
spectrum shows a weak feature, which is caused almost entirely by [\FeII] lines.
If a weak underlying continuum was removed from the data the discrepancy would
not be as large. Another possibility is poorly known atomic data for these
lines. This may also affect the emission near 9000\,\AA, which our spectra only
partially reproduce, while we have no suitable candidate for the emission near
9500\,\AA. The emissions near 9000 and 9500\,\AA\ were also seen in SNe\,1991bg
and 1999by, while the one near 8100\,\AA\ was not \citep[Fig.
\ref{fig:10lp_91bg}, also][]{taub13}. 

One scenario in which central oxygen emission has been predicted is that of the
deflagration of a Chandrasekhar-mass carbon-oxygen white dwarf. In the slow
burning of a deflagration wave mixing can drag unburned material to the inner
parts of the ejecta. This configuration is expected to produce strong central
[\OI] emission \citep{kozma05}.  However, the low central density and the
implied low ejected mass that are inferred from the comparison with SN\,1999bg
and the nebular spectral modelling both suggest that SN\,2010lp was itself a
sub-\MCh\ SN\,Ia. 

A number of possible scenarios may then be invoked. Accretion of helium from a
companion on the surface of a sub-\MCh\ mass white dwarf can cause the surface
of the white dwarf to ignite \citep{livne95}. If the ensuing shock wave can
propagate and focus at the centre of the star, the white dwarf can explode via a
central detonation \citep{Shen14}. Given the requirement of central ignition,
this scenario is not likely to lead to two distinct blobs of unburned material. 

The violent merger of two sub-\MCh\ mass white dwarfs can also lead to ignition
and explosion, as discussed above. If the two white dwarfs are of sufficiently
different mass, as in the case presented by \citet{kromer13}, low-velocity
oxygen may be the result of incomplete burning of some part of the less massive
white dwarf that is disrupted in the merger process. \citet{kromer13} present a
simulation of a merger of two white dwarfs of mass 0.9 and 0.76\,\Msun,
respectively, and show that as much as 0.5\,\Msun\ of oxygen from the disrupted
secondary are left at $v \lsim 2000$\kms. This is much more than we find in our
simulations of the nebular spectrum of SN\,2010lp, and it has a spherical
distribution, which would not be compatible with the spectrum of SN\,2010lp, but
it is at least one step in the required direction. Compared to their model,
SN\,2010lp produced much less oxygen, and it ejected it in two blobs. They also
used a larger value of the reddening than we find here, and therefore had to use
a 0.9\,\Msun\ white dwarf as the primary of the system in order to synthesize
enough \Nifs\  (0.18\,\Msun). With the smaller reddening determined here the
amount of \Nifs\ necessary to fit the light curve of SN\,2010lp would be less.
We estimate here that $\approx 0.08$\,\Msun of \Nifs\ were sufficient to
energise the nebular spectrum of SN\,2010lp, while for SN\,1991bg
\citet{mazz91bgneb} estimated 0.06\,\Msun. Some merger scenario with slightly
different masses and mass ratios might produce the required mass and
distribution of oxygen. 

A third possibility is that the two blobs mark part of the material that
survives the explosion following the impact (collision) of two sub-\MCh\ white
dwarfs, as in the models of \citet{kushnir13}. Although the original model of
white dwarfs collision envisages the creation of two central blobs of \Nifs\, it
may not be excluded that some particular configuration, or range of white dwarf
masses, may lead to an outcome similar to what is diagnosed in SN\,2010lp. The
question in this case might be, what would then set these inner regions into
expansion? 

Finally, another possible way to place material at low velocity is for it to be
companion material stripped off and swept up by the impact with the SN ejecta,
such as is expected for a hydrogen-rich companion \citep{marietta2000}. Could
oxygen lost from a double-degenerate system during merging lead to such a
configuration? What would cause two distinct blobs?

\section{Conclusions}
\label{sec:concl}

Synthetic spectra have been computed for the nebular spectrum of the
sub-luminous, peculiar SN\,Ia 2010lp. The overall spectrum can be reproduced
reasonably well if a sub-\MCh\ explosion model is used, as in SN\,1991bg. In
SN\,1991bg the presence of low-velocity \Nifs\ is consistent with a scenario
that involves the violent merger of two sub-\MCh\ mass white dwarfs, although
the mass of \Nifs\ synthesised should be less than in the models that have been
produced so far. 

In the case of SN\,2010lp, the observed low-velocity, double peaked [\OI]
emission can be reproduced replacing the central \Nifs\ with oxygen.  The amount
of oxygen that is required to produce the observed emission is actually small
($\sim 0.05$\,\Msun), and the emission is caused by two oxygen-rich blobs,
moving in opposite directions. 

If two low-mass white dwarfs merged or collided to give rise to the SN, their
respective central regions should suffer very limited burning in order to
reproduce the observations of SN\,2010lp. The lack of corresponding low-velocity
carbon emission lines suggests that at least carbon from the progenitor did burn
to oxygen. 

Some merger models predict the presence of unburned oxygen at low velocities,
but no published model predicts the formation of separate oxygen blobs. Such
blobs may possibly be observed in more detailed three-dimensional hydrodynamical
simulations, covering perhaps a different space of parameters (white dwarf
masses, mass ratio), or may require the development of a completely new
scenario.

\section*{Acknowledgments}
M. Stritzinger is supported by the VILLUM FONDEN (grant number 28021) and the
Independent Research Fund Denmark (IRFD; 8021-00170B). C. Ashall is supported by
NASA grant 80NSSC19K1717 and NSF grants AST-1920392 and AST-1911074.  The
maximum-light spectrum of SN\,2010lp was obtained with the Gemini North
telescope as part of program identification GN-2010B-Q-67 (PI M. Stritzinger).
The authors wish to recognize and acknowledge the very significant cultural role
and reverence that the summit of Maunakea has always had within the indigenous
Hawaiian community. We are most fortunate to have the opportunity to conduct
observations from this mountain. \\
We wish to thank the anonymous referee for a fair, competent report.

\section*{Data availability}
The photometric and spectroscopic data presented in this article are
publicly available via the Weizmann Interactive Supernova Data
Repository,  at https://wiserep.weizmann.ac.il.

\end{document}